\documentclass[a4paper,11pt]{article}
\pdfoutput=1 
\usepackage{jheppub} 
\usepackage[T1]{fontenc} 
\usepackage{multirow}
\usepackage{color}
\usepackage{ulem}

\usepackage{xcolor}
\usepackage{threeparttable}
\usepackage{multirow}
\usepackage{subfigure}
\usepackage{float}
\usepackage{caption}

\title{\boldmath Study of the processes $\chi_{cJ} \to \Xi^- \bar{\Xi}^+$ and $\Xi^0 \bar{\Xi}^0$}

\collaborationImg{\includegraphics[width=0.15\textwidth]{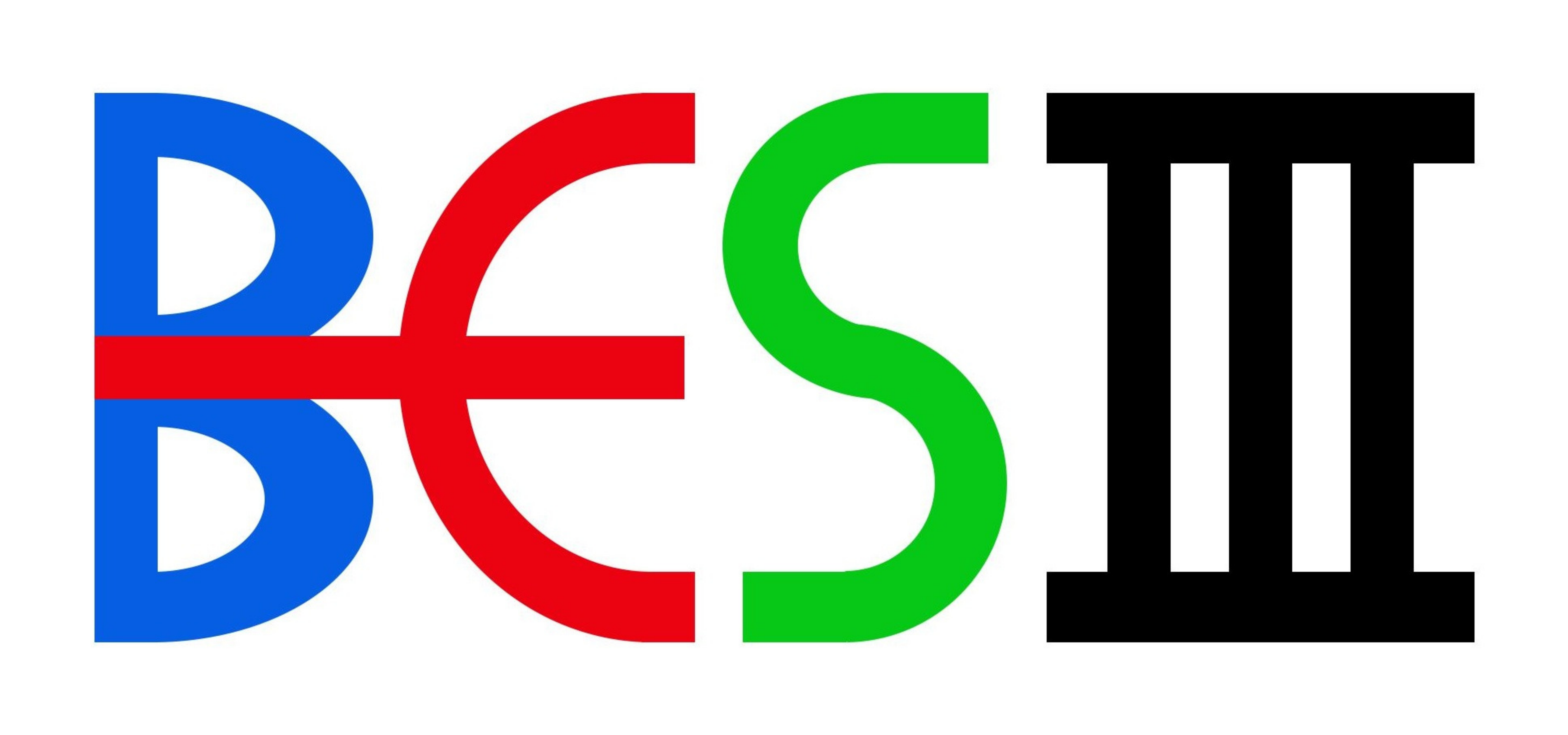}}

\collaboration{The BESIII Collaboration}

\author{
	M.~Ablikim$^{1}$, M.~N.~Achasov$^{10,b}$, P.~Adlarson$^{68}$, S. ~Ahmed$^{14}$, M.~Albrecht$^{4}$, R.~Aliberti$^{28}$, A.~Amoroso$^{67A,67C}$, M.~R.~An$^{32}$, Q.~An$^{64,50}$, X.~H.~Bai$^{58}$
% ...
}
\affiliation{
	$^{1}$ Institute of High Energy Physics, Beijing 100049, People's Republic of China\\
% ...
$^{4}$ Bochum Ruhr-University, D-44780 Bochum, Germany\\
% ...
$^{10}$ G.I. Budker Institute of Nuclear Physics SB RAS (BINP), Novosibirsk 630090, Russia\\
% ...
$^{14}$ Helmholtz Institute Mainz, Staudinger Weg 18, D-55099 Mainz, Germany\\
% ...
$^{28}$ Johannes Gutenberg University of Mainz, Johann-Joachim-Becher-Weg 45, D-55099 Mainz, Germany\\
% ...
$^{32}$ Liaoning Normal University, Dalian 116029, People's Republic of China\\
% ...
$^{50}$ State Key Laboratory of Particle Detection and Electronics, Beijing 100049, Hefei 230026, People's Republic of China\\
% ...
$^{58}$ University of Jinan, Jinan 250022, People's Republic of China\\
% ...
$^{64}$ University of Science and Technology of China, Hefei 230026, People's Republic of China\\
% ...
$^{67}$ University of Turin and INFN, (A)University of Turin, I-10125, Turin, Italy; (B)University of Eastern Piedmont, I-15121, Alessandria, Italy; (C)INFN, I-10125, Turin, Italy\\
	$^{68}$ Uppsala University, Box 516, SE-75120 Uppsala, Sweden\\
% ...
$^{b}$ Also at the Novosibirsk State University, Novosibirsk, 630090, Russia\\
% ...
}

\emailAdd{besiii-publications@ihep.ac.cn}

\abstract{Using $448.1\times 10 ^6$ $\psi(3686)$ decays collected with the BESIII detector at the BEPCII $e^+e^-$ storage rings, the branching fractions and angular distributions of the decays $\chi_{cJ}\to\Xi^-\bar{\Xi}^+$ and $\Xi^0\bar{\Xi}^0$ $(J = 0, 1, 2)$ are measured based on a partial-reconstruction technique. The decays $\chi_{c1}\to\Xi^0\bar{\Xi}^0$ and $\chi_{c2}\to\Xi^0\bar{\Xi}^0$ are observed for the first time with statistical significances of 7$\sigma$ and 15$\sigma$, respectively. The results of this analysis are in good agreement with previous measurements and have  significantly improved precision.}

\begin{document} 
\maketitle
\flushbottom

\section{Introduction}
\label{sec:intro}

Experimental studies of hadronic two-body decays of charmonium states, in particular those into baryon--anti-baryon pairs, are essential for testing perturbative Quantum Chromodynamics (QCD), e.g. the helicity selection rule (HSR), which prohibits $\chi_{c0}$ decays to baryon--anti-baryon pairs \cite{hsrpaper}. However, it has long been known that this rule is strongly violated. The measured branching fractions for the $\chi_{c0}$ meson decaying into a number of baryon--anti-baryon final states do not vanish \cite{exp1, exp2, exp3, bes2_xixibar, exp4, exp5, exp6, exp7}, which indicates substantial contributions due to finite quark masses. These observations have stimulated many theoretical efforts \cite{theory1, theory2, theory3, theory4, theory5}. In refs. \cite{theory4, theory5}, it was proposed that intermediate-meson loop transitions can serve as a soft mechanism in charmonium decays, and such a long-distance interaction can bypass the Okubo-Zweig-Iizuka rule and lead to a violation of the HSR. In the framework of effective Lagrangians for hadron interactions, ref. \cite{theory1} performed a quantitative study of this mechanism investigating the role of charmed hadron loops in $\chi_{c0}$ decays to baryon anti-baryon pairs. However, considerable discrepancies have been seen between theoretical predictions and experimental results. This could be a sign that SU(4) flavor symmetry is badly broken. In this case, more precise measurements and more decay channels are needed to understand the mechanisms of HSR violation.

The importance of the color-octet mechanism (COM) in $\chi_{cJ}$ decays was pointed out more than two decades ago, and this concept was invoked in many theoretical predictions of exclusive two-body decays of the $\chi_{cJ}$ meson~\cite{com1, com2, com3}. In this mechanism, the $\chi_{cJ}$ state is treated as more than just a pure quark-antiquark state and the octet operators are included in the transition matrix elements to a particular final state to calculate the exclusive two-body decay rates \cite{com2}. In the case of $\chi_{cJ}$ decays to meson pairs, the COM predictions are in agreement with experiments, while for baryon pairs, the COM cannot give a satisfactory prediction, e.g. ref.~\cite{com3} has used the COM to explain the partial decay widths of $\chi_{cJ} \to p\bar{p}$ processes and made predictions for the processes $\chi_{cJ} \to \Lambda\bar{\Lambda}$, $\Sigma\bar{\Sigma}$ and $\Xi\bar{\Xi}$. The predictions show that the decay widths decrease roughly with increasing baryon masses. However, the predictions are  significantly lower than the experimental results. Therefore more decay modes of the $\chi_{cJ}$ mesons are desired as inputs to further study the contributions of the COM.

The decay mechanisms of the $\chi_{cJ}$ meson can be studied by measuring the angular distribution of the final state particles. In the helicity frame, the angular distribution of a spin-1/2 baryon from charmonium decays takes the form~\cite{helicity_amp}
\begin{equation}\label{fad}
\frac{dN}{d\cos\theta}\propto(1 + \alpha\cos^2 \theta),
\end{equation}
where $\theta$ is the helicity angle of the final state baryon with respect to the initial state charmonium and $\alpha$ is a constant indicating the level of polarization of the initial state charmonium, which has been widely studied both in theory and experiment~\cite{adpaper1, adpaper2, singletag0}.

In the Standard Quark Model, the $\chi_{cJ}$ mesons are $c\bar{c}$ states in a $L = 1$ configuration and cannot be produced directly in $e^+e^-$ annihilation. However, the large branching fractions of the decays $\psi(3686)\to\gamma\chi_{cJ}$ provides very clean environments for the investigation of the decay mechanism of the $\chi_{cJ}$ mesons. More than ten years ago BESII and CLEO-c measured the branching fractions of the processes $\chi_{cJ} \to \Xi^-\bar{\Xi}^+$ and $\Xi^0\bar{\Xi}^0$~\cite{exp3, bes2_xixibar}. Due to the limited sample sizes, no significant signals were observed for the $\chi_{c1} \to \Xi^0\bar{\Xi}^0$ and $\chi_{c2} \to \Xi^0\bar{\Xi}^0$ processes, and the other processes were measured with large uncertainties.

In this paper, we present a measurement of branching fractions and angular distributions for the processes $\chi_{cJ} \to \Xi^-\bar{\Xi}^+$ and $\Xi^0\bar{\Xi}^0$. The data sample used in this analysis corresponds to a total of $(448.1 \pm 2.9) \times 10^6$ $\psi(3686)$ decays~\cite{DATASET} collected with the BESIII detector~\cite{BESIII} at BEPCII~\cite{BEPCII}.

\section{BESIII Detector and Monte Carlo Simulation}
The BESIII detector~\cite{BESIII} records symmetric $e^+e^-$ collisions provided by the BEPCII storage ring~\cite{BEPCII}, which operates with a peak luminosity of 1$\times$10$^{33}$~cm$^{-2}$s$^{-1}$ in the center-of-mass energy range from 2.0 to 4.9 GeV. BESIII has collected large data samples in this energy region~\cite{BESIII_DATASET}. The cylindrical core of the BESIII detector covers 93\% of the full solid angle and consists of a helium-based multilayer drift chamber (MDC), a plastic scintillator time-of-flight system (TOF), and a CsI(Tl) electromagnetic calorimeter (EMC), which are all enclosed in a superconducting solenoidal magnet that provides a 1.0~T magnetic field. The solenoid is supported by an octagonal flux return yoke with resistive plate counter muon identifier modules interleaved with steel. The momentum resolution for charged particles at 1~GeV/$c$ is 0.5\%, and the d$E$/d$x$ resolution for electrons from Bhabha scattering is 6\%. The EMC measures photon energies with a resolution of 2.5\% (5\%) at 1 GeV in the barrel (end-cap) region. The time resolution of the TOF barrel part is 68~ps, while that in the end-cap part is 110~ps.

Simulated data samples produced with a {\sc geant4}-based~\cite{GEANT4} Monte Carlo (MC) package, which includes the geometric description of the BESIII detector and the detector response, are used to determine detection efficiencies and to estimate backgrounds. The simulation models the beam energy spread and initial-state radiation (ISR) in the $e^+e^-$ annihilation using the generator {\sc kkmc}~\cite{KKMC}. The inclusive MC sample includes the production of the $\psi(3686)$ resonance, the ISR production of the $J/\psi$ meson, and the continuum processes incorporated in {\sc kkmc}~\cite{KKMC}. Known decay modes are modeled with {\sc evtgen}~\cite{EVTGEN, evtgen2} using branching fractions taken from the Particle Data Group (PDG)~\cite{PDG}, and the remaining unknown charmonium decays are modeled with {\sc lundcharm}~\cite{LUNDCHARM, lundcharm2}. Final-state radiation (FSR) from charged final state particles is included using the {\sc photos} package~\cite{PHOTOS}. To determine the detection efficiency, exclusive MC samples are generated for each signal process. The decay of $\psi(3686)\to\gamma\chi_{cJ}$ is generated by taking the angular distribution from refs.~\cite{E1}, where the helicity angle $\theta_{\gamma}$ of the radiative photon is distributed according to $1 + \alpha'\cos^2\theta_{\gamma}$, and $\alpha' = 1$, $-1/3$, $1/13$ for $J = 0, 1$ and $2$, respectively. It has been argued that the electric dipole (E1) dominates the decay of $\psi(3686)\to\gamma\chi_{cJ}$, which was confirmed by the BESIII measurement~\cite{M2E3}. Hence the higher-order multipole amplitudes are neglected in this analysis. The decays of $\chi_{cJ}\to\Xi^-\bar{\Xi}^+$ and $\Xi^0\bar{\Xi}^0$ are generated with the measured angular distributions, where the helicity angle $\theta_{b}$ of the outgoing baryon satisfies the angular distribution $1 + \alpha\cos^2\theta_{b}$. Note that the decay of a scalar meson is isotropic, so $\alpha = 0$ for the decays of the $\chi_{c0}$ meson. The decays of the $\Xi$ baryons as well as their anti-particles are inclusively simulated via \textsc{evtgen}~\cite{EVTGEN, evtgen2}.

\section{Event Selection}
Since the full reconstruction method suffers from low selection efficiency, a partial reconstruction technique~\cite{cleoxi, singletag,  BESIII:2021aer} is used in this analysis. The radiative photon from the decay $\psi(3686)\to\gamma\chi_{cJ}$ is reconstructed to infer the presence of a $\chi_{cJ}$ meson. The $\Xi^-$ ($\Xi^0$) baryon is reconstructed by the decay to $\Lambda\pi^-$ ($\Lambda\pi^0$)  with the subsequent decay $\Lambda\to p\pi^-$ ($\pi^0\to\gamma\gamma$), and the presence of the anti-baryon $\bar{\Xi}^+$ ($\bar{\Xi}^0$) is inferred from the invariant mass of the system recoiling against the reconstructed $\gamma\Xi$ system (unless otherwise noted, the charge conjugated state is implicitly included throughout the paper).

Tracks of charged particles detected in the MDC are required to satisfy $|\cos\theta| < 0.93$, where the polar-angle $\theta$ is defined with respect to the $z$-axis. The particle identification (PID) for  charged particles combines measurements of the energy deposited in the MDC and the time of flight by the TOF to form likelihoods ${\cal L}(h)$ $(h = p, K, \pi)$ for each hadron $h$ hypothesis. Tracks are identified as protons if the proton hypothesis has the greatest likelihood $({\cal L}(p) > {\cal L}(K)$ and ${\cal L}(p) > {\cal L}(\pi))$, while charged kaons and pions are identified by comparing the likelihoods for the kaon and pion hypotheses, ${\cal L}(K) > {\cal L}(\pi)$ and ${\cal L}(\pi) > {\cal L}(K)$, respectively. Events with two (one) negatively charged pions and one proton are kept for further analysis of the $\chi_{cJ}\to\Xi^-\bar{\Xi}^+$ ($\Xi^0\bar{\Xi}^0$) process.

\begin{figure}[tbp]
	\centering
	\includegraphics[width=0.6\textwidth]{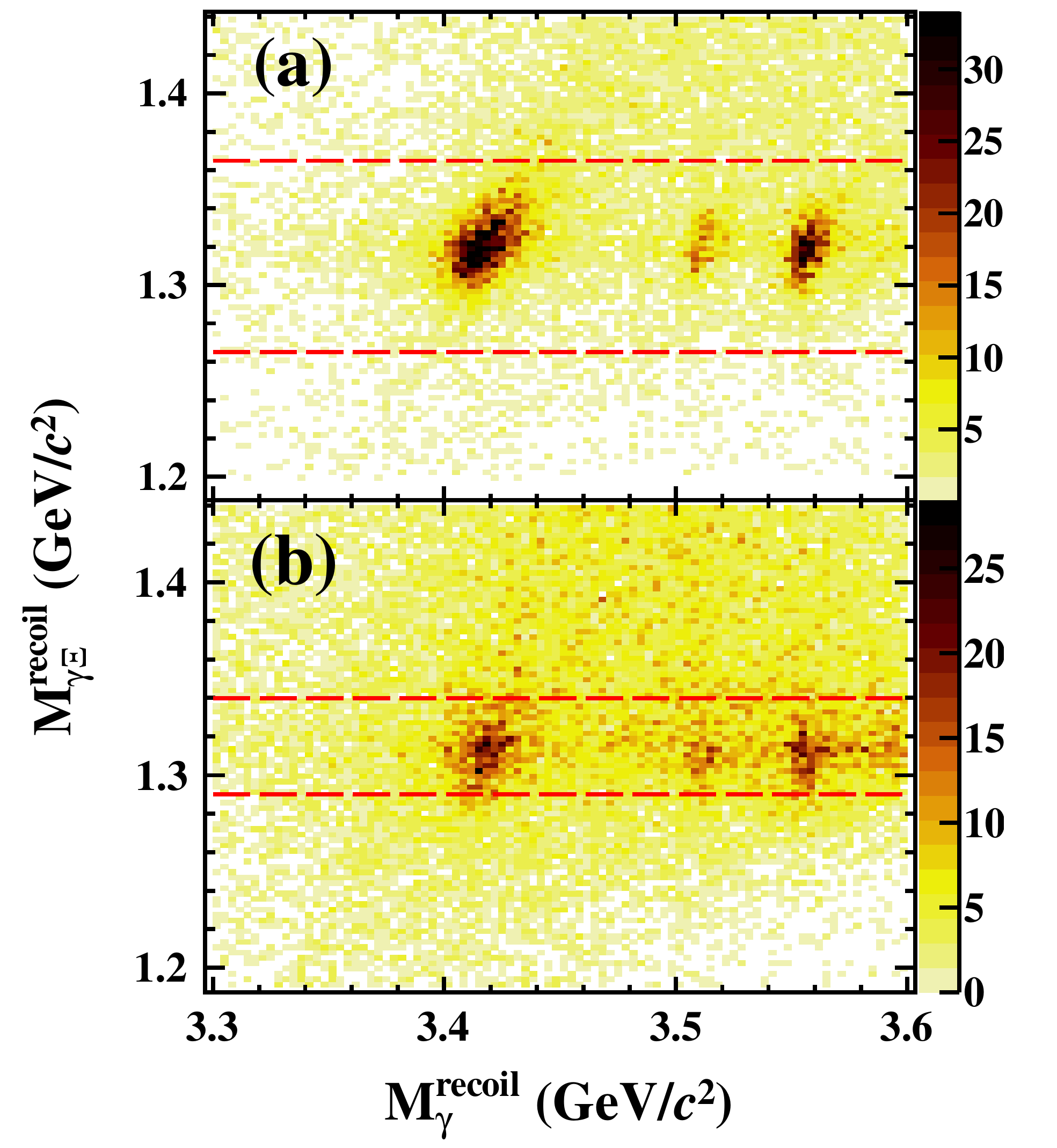}
	\caption{\label{scatterplot} 2D distributions of $M^{\rm recoil}_{\gamma\Xi}$ versus $M^{\rm recoil}_{\gamma}$ from data for the (a) $\chi_{cJ}\to\Xi^-\bar{\Xi}^+$ and (b) $\chi_{cJ}\to\Xi^0\bar{\Xi}^0$ processes. The dashed lines indicate the signal region.}
\end{figure}

Photon candidates are identified using showers in the EMC. The deposited energy of each shower must be greater than 25 MeV in the barrel region $(|\cos\theta| < 0.80)$ and greater than 50 MeV in the end-cap region $(0.86 < |\cos\theta| < 0.92)$. To exclude showers that originate from tracks of charged particles, the angle between the position of each shower in the EMC and the closest extrapolated track  must be greater than 10$^{\circ}$. To suppress electronic noise and showers unrelated to the event, the difference between the EMC time and event start time has to be between 0 and 700~ns. At least one (three) photon candidates are required for further study of the $\chi_{cJ}\to\Xi^-\bar{\Xi}^+$ ($\Xi^0\bar{\Xi}^0$) process.

The reconstruction of $\Lambda$ and $\Xi^-$ baryons follow the technique explained in refs.~\cite{ee2xixibar, BESIII:2021cvv, BESIII:2021ccp}. Briefly, to reconstruct $\Lambda$ candidates, a secondary vertex fit~\cite{vtxfit} is applied to all $p\pi^-$ combinations, and their invariant masses are required to be within 5 MeV/$c^{2}$ of the known $\Lambda$ mass from the PDG~\cite{PDG}. The decay length of the $\Lambda$ candidate, i.e. the distance between its production and decay position, is required to be greater than zero. Similarly, $\Xi^{-}$ candidates are reconstructed with $\Lambda\pi^{-}$ combinations, and their invariant masses are required to be within 10 MeV/$c^{2}$ of the known $\Xi^{-}$ mass~\cite{PDG}. The decay length of the $\Xi^{-}$ candidate is also required to be greater than zero.

To reconstruct the $\pi^0$ meson from the $\Xi^0\to\Lambda\pi^0$ decay, a one-constraint (1C) kinematic fit is applied to all $\gamma\gamma$ combinations using the $\pi^0\to\gamma\gamma$ hypothesis. The requirement $\chi^{2}_{\rm 1C} < 20$ is employed to suppress the non-$\pi^0$ backgrounds~\cite{singletag0}. The candidate with the minimum value of $|M_{\gamma\gamma\Lambda} - m_{\Xi^0}|$ among all $\gamma\gamma$ combinations is selected, where $M_{\gamma\gamma\Lambda}$ is the invariant mass of the $\gamma\gamma\Lambda$ system, $m_{\Xi^0}$ is the known $\Xi^0$ mass~\cite{PDG}, and the value of $|M_{\gamma\gamma\Lambda} - m_{\Xi^0}|$ is required to be less than 10 MeV/$c^{2}$. The selected $\pi^0$ candidate together with the $\Lambda$ candidate are combined to form the $\Xi^0$ candidate.

\begin{figure}[tbp]
	\centering
	\includegraphics[width=0.7\textwidth]{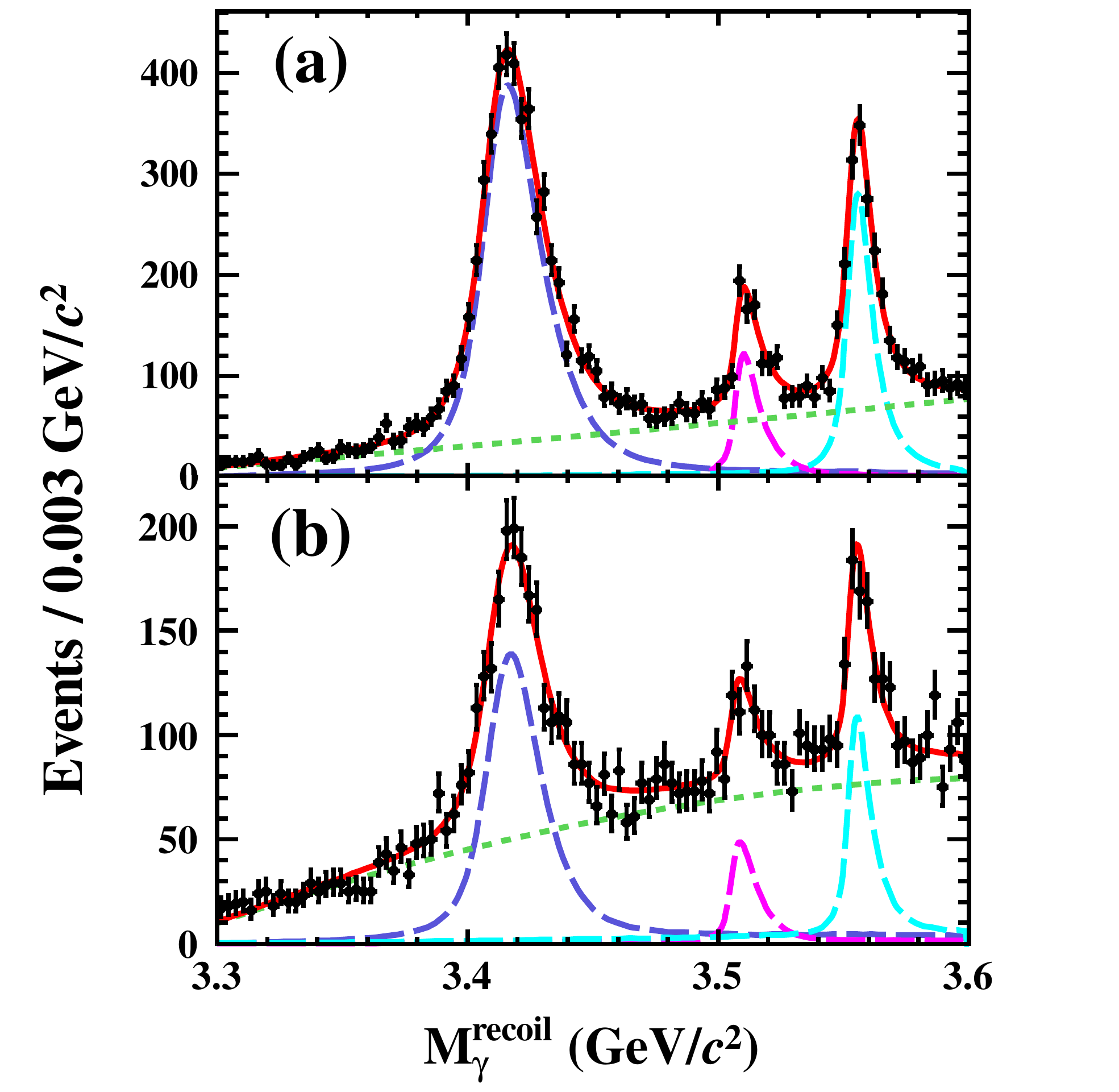}
	\caption{\label{fitting} Fit of the $M^{\rm recoil}_{\gamma}$ distributions for the (a) $\chi_{cJ}\to\Xi^-\bar{\Xi}^+$ and (b) $\chi_{cJ}\to\Xi^0\bar{\Xi}^0$ processes. Dots with error bars are data, the solid lines show the fit results, the long dashed lines represent the signal contributions, and the short dashed lines represent the smooth backgrounds.}
\end{figure}

The radiative photon from  the $\psi(3686)\to\gamma\chi_{cJ}$ decay is selected from the remaining photon candidates. The candidate with the minimum value of $|M_{\gamma\Xi}^{\rm recoil} - m_{\Xi}|$ is selected for the $\chi_{cJ}\to\Xi^-\bar{\Xi}^+$ and $\Xi^0\bar{\Xi}^0$ processes. Here $m_{\Xi}$ is the known $\Xi$ mass, and  $M_{\gamma\Xi}^{\rm recoil}$ is the recoiling mass of the $\gamma\Xi$ system, which for signal events is the invariant mass of the $\bar{\Xi}$  anti-baryon. The recoiling mass of $\gamma\Xi$ system is required to be within 50 (25)~ MeV/$c^{2}$ for the $\chi_{cJ}\to\Xi^-\bar{\Xi}^+$ ($\Xi^0\bar{\Xi}^0$) decay.

The $\chi_{cJ}$ candidate can be inferred from the system recoiling against the selected radiative photon, i.e. $M_{\gamma}^{\rm recoil}$. Figure~\ref{scatterplot} shows the two-dimensional (2D) distributions of $M^{\rm recoil}_{\gamma\Xi}$ versus $M^{\rm recoil}_{\gamma}$ for data. Clear accumulations around the known $\Xi$ and $\chi_{cJ}$ masses can be seen. Potential sources of background are investigated by studying the generic MC samples after imposing the signal-selection criteria with an event-type analysis tool, TopoAna~\cite{topoana}. It is found that the decay $\psi(3686)\to\pi^{0(-)}\pi^{0(+)} J/\psi$ ($J/\psi\to\Lambda\bar{\Lambda}$) is the dominant background process, and is distributed smoothly throughout the region of interest.

\begin{figure}[tbp]
	\centering
	\includegraphics[width=0.48\textwidth]{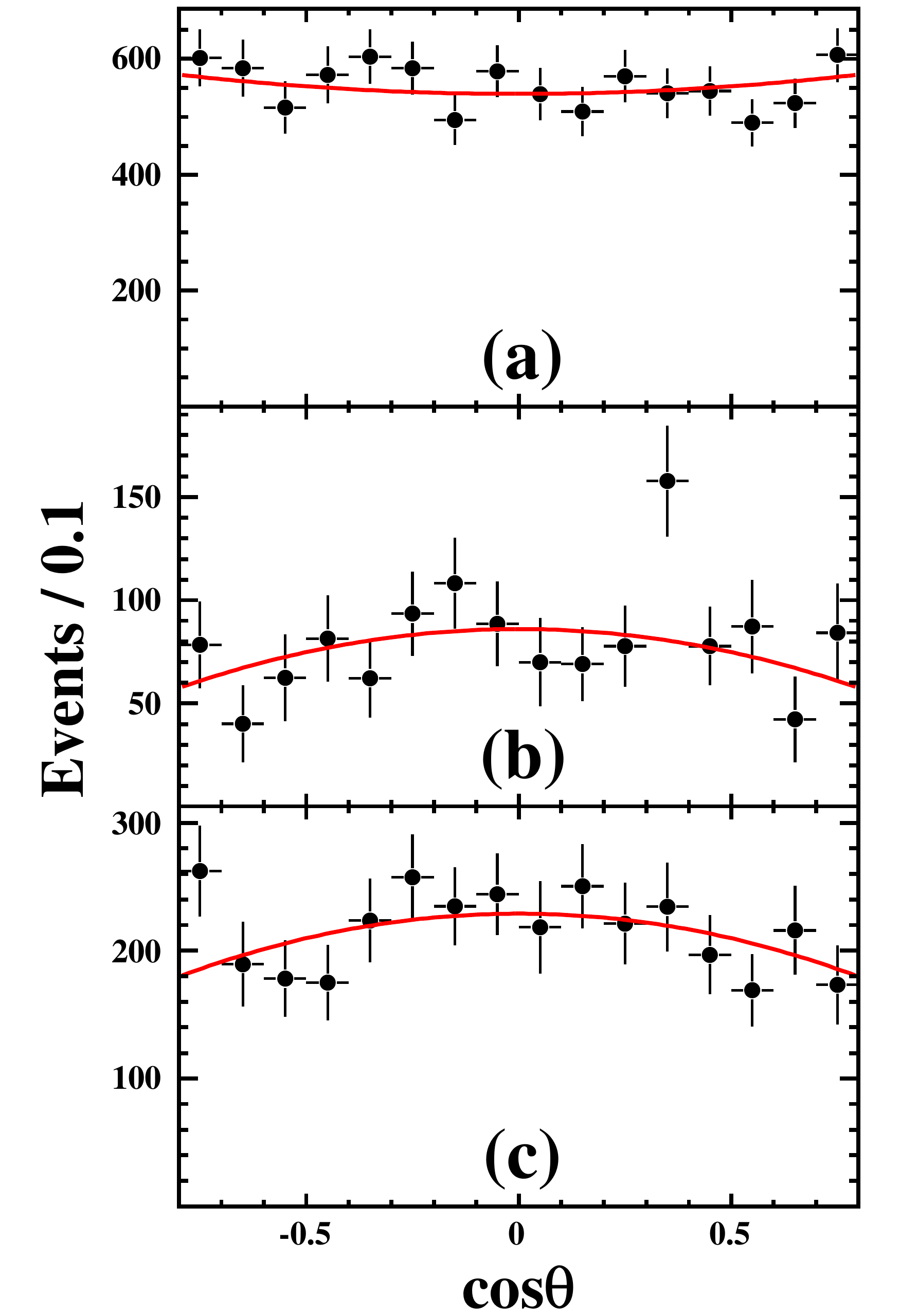}
	\hfill
	\includegraphics[width=0.48\textwidth]{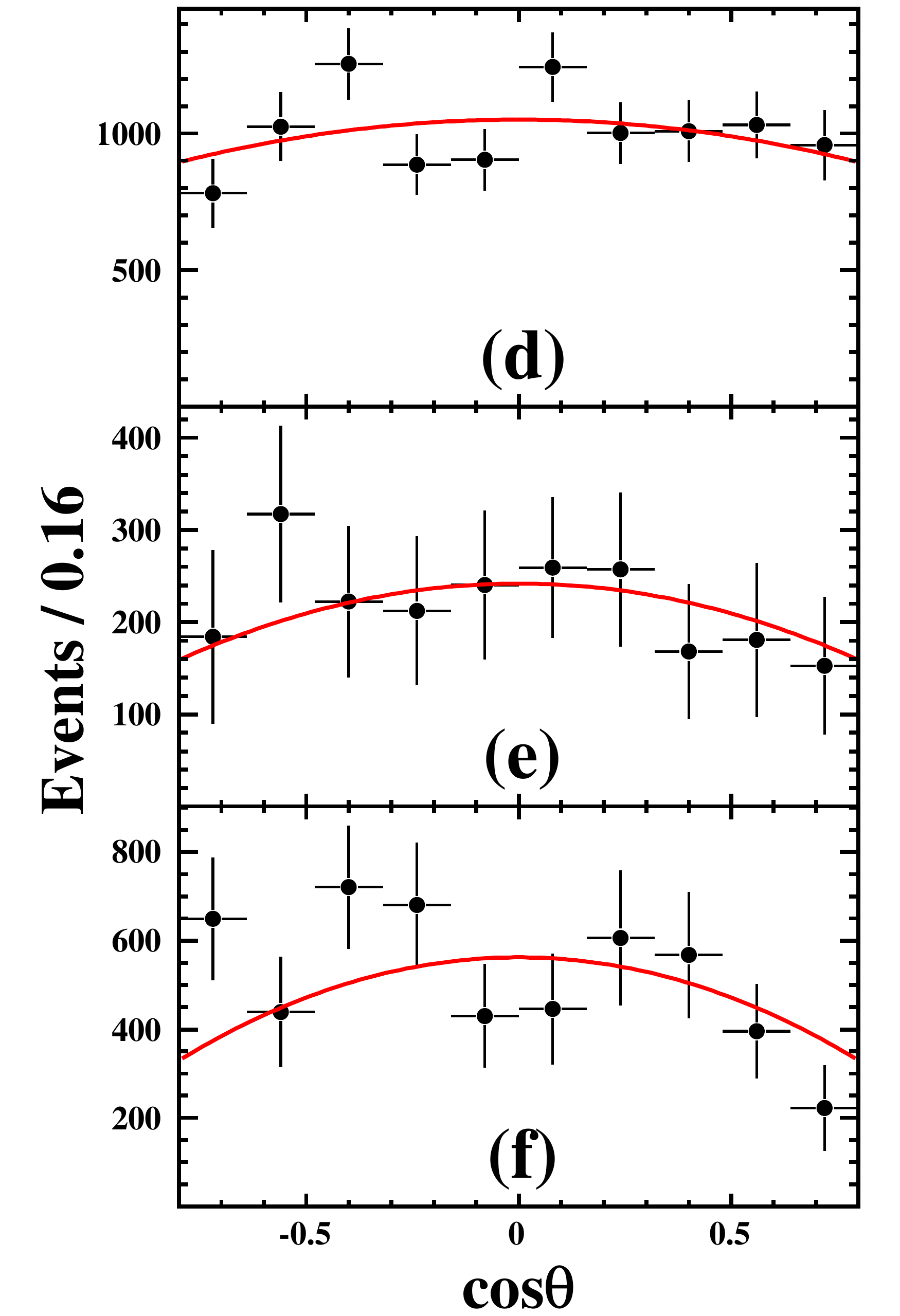}
	\caption{\label{fit_ad} Angular distributions for $\chi_{cJ}\to\Xi^-\bar{\Xi}^+$ processes with (a) $J = 0$, (b) $J = 1$, (c) $J = 2$ and $\chi_{cJ}\to\Xi^0\bar{\Xi}^0$ processes with (d) $J = 0$, (e) $J = 1$, (f) $J = 2$. Dots with error bars are efficiency-corrected data and the solid lines show the fit results.}
\end{figure}

\section{Determination of Branching Fractions and Decay Angular Distributions}
\subsection{Determination of Branching Fractions}
The signal yield for each process is determined by performing an unbinned maximum likelihood fit to the $M^{\rm recoil}_{\gamma}$ spectrum in the range between 3.3 and 3.6~GeV/$c^{2}$. The signal shape for each process is represented by the individual MC simulated shape convolved with a Gaussian function to compensate for the mass resolution difference between data and MC, and the parameters of the Gaussian function are left free in the fit. The background is described by a second-order polynomial. Figure~\ref{fitting} shows the $M^{\rm recoil}_{\gamma}$ distributions and the fit results for each process.

The branching fractions are calculated as
\begin{equation}
\mathcal{B}(\chi_{cJ}\to\Xi\bar{\Xi})=\frac{N_{\rm obs}}{N_{\psi(3686)}\cdot\mathcal{B}(\psi(3686)\to\gamma\chi_{cJ})\cdot\epsilon},
\end{equation}
where $N_{\rm obs}$ is the number of observed signal events, $\epsilon$ is the detection efficiency of the single baryon-tag technique, including the branching fractions for the subsequent decays of the $\Lambda$ and $\Xi$ baryons, $N_{\psi(3686)}$ is the total number of $\psi(3686)$ decays~\cite{DATASET} and ${\cal{B}}(\psi(3686)\to\gamma\chi_{cJ})$ is the branching fraction of the $\psi(3686)\to\gamma\chi_{cJ}$ decay~\cite{PDG}.

\subsection{Decay Angular Distributions}
The angular distribution for each process is studied using a binned method~\cite{binbybin}. The data satisfying all selection criteria are divided into 16 (10) intervals in $\cos \theta$ from $-0.8$ to 0.8 for the $\chi_{cJ}\to\Xi^-\bar{\Xi}^+$ ($\Xi^0\bar{\Xi}^0$) process. The signal yield in each $\cos \theta$ interval is determined by a fit to the $M_{\gamma}^{\rm recoil}$ spectrum. A bin-by-bin efficiency correction is applied to the signal yield, where the corresponding efficiency is obtained from MC simulation. Since the decay of a scalar meson is isotropic, the measurements of the angular distribution in $\chi_{c0}$ decays are regarded as a systematic check for the other measurements. The efficiency-corrected angular distributions are fitted by eq.(\ref{fad}) as shown in figure \ref{fit_ad}. The fit results for $\chi_{c0}$ are consistent with theoretical expectations~\cite{helicity_amp}.

\begin{table}[bpt]
	\centering
	\begin{tabular}{|ccccc|}
		\hline
		Channel & $N_{\rm obs}$ & $\epsilon$ (\%) & $\alpha$ & ${\cal B}$ ($\times 10^{-4}$) \\
		\hline
		\multirow{2}{*}{$\chi_{c0}\to$ $\begin{matrix} ~\Xi^-\bar{\Xi}^+\\ \Xi^0 ~ \bar{\Xi}^0 \end{matrix}$} & 4932 $\pm$ 92 & 25.4 & ~~ 0.09 $\pm$ 0.11 $\pm$ 0.17 & 4.43 $\pm$ 0.08 $\pm$ 0.18 \\
		& 1741 $\pm$ 71 & ~~8.5 & $-$0.23 $\pm$ 0.19 $\pm$ 0.36 & 4.67 $\pm$ 0.19 $\pm$ 0.26 \\
		\multirow{2}{*}{$\chi_{c1}\to$ $\begin{matrix} ~\Xi^-\bar{\Xi}^+\\ \Xi^0 ~ \bar{\Xi}^0 \end{matrix}$} & ~~692 $\pm$ 44 & 27.3 & $-$0.52 $\pm$ 0.29 $\pm$ 0.48 & 0.58 $\pm$ 0.04 $\pm$ 0.05 \\
		& ~~325 $\pm$ 49 & ~~9.9 & $-$0.54 $\pm$ 0.52 $\pm$ 0.43 & 0.75 $\pm$ 0.11 $\pm$ 0.06 \\
		\multirow{2}{*}{$\chi_{c2}\to$ $\begin{matrix} ~\Xi^-\bar{\Xi}^+\\ \Xi^0 ~ \bar{\Xi}^0 \end{matrix}$} & 1691 $\pm$ 66 & 27.6 & $-$0.34 $\pm$ 0.18 $\pm$ 0.30 & 1.44 $\pm$ 0.06 $\pm$ 0.11 \\
		& ~~804 $\pm$ 67 & 10.3 & $-$0.65 $\pm$ 0.31 $\pm$ 0.22 & 1.83 $\pm$ 0.15 $\pm$ 0.16 \\
		\hline
	\end{tabular}
	\caption{\label{sum} The measured branching fractions $\cal B$ and decay parameters $\alpha$ for the  $\chi_{cJ}\to\Xi^-\bar{\Xi}^+$ and $\chi_{cJ}\to\Xi^0\bar{\Xi}^0$ processes, where $N_{\rm obs}$ is the number of signal events and $\epsilon$ is the detection efficiency. The first uncertainty is statistical and the second is systematic.}
\end{table}

\begin{table}[bpt]
	\centering
	\begin{tabular}{|cccc|}
		\hline
		& $\chi_{c0}$ & $\chi_{c1}$ & $\chi_{c2}$ \\
		\hline
		Ratio & 0.95 $\pm$ 0.04 $\pm$ 0.06 & 0.77 $\pm$ 0.12 $\pm$ 0.08 & 0.79 $\pm$ 0.07 $\pm$ 0.09 \\
		\hline
	\end{tabular}
	\caption{\label{ratio} The ratio between the branching fractions of the  $\chi_{cJ}\to\Xi^-\bar{\Xi}^+$ and $\chi_{cJ}\to\Xi^0\bar{\Xi}^0$ processes. The first uncertainty is statistical and the second is systematic. Systematic uncertainties arising from the same source cancel when calculating the ratio.}
\end{table}

Table~\ref{sum} summarizes the measured branching fractions and $\alpha$ parameters. To test isospin symmetry, the ratios between the branching fractions of the processes $\chi_{cJ}\to\Xi^-\bar{\Xi}^+$ and $\chi_{cJ}\to\Xi^0\bar{\Xi}^0$ are calculated and summarized in table~\ref{ratio}. The single-baryon tag method leads to a double counting effect for the $\Xi\bar{\Xi}$ final state,  which is taken into account in the calculation of the statistical uncertainties based on the study of MC simulation~\cite{doublecounting}. In this analysis, the double counting ratio is about 20\% (6.5\%) for the $\chi_{cJ}\to\Xi^-\bar{\Xi}^+$ ($\Xi^0\bar{\Xi}^0$) process. This effect does not affect the central value of the final result but it does affect the statistical uncertainty. If the uncertainty is determined by the fit, the relative statistical uncertainty is underestimated by 8\% (3\%) approximately.

\section{Systematic Uncertainty}

\begin{table}[bpt]
	\centering
	\begin{tabular}{|ccccccccc|}
		\hline
		& \multicolumn{3}{c}{$\chi_{cJ}\to\Xi^-\bar{\Xi}^+$} && \multicolumn{3}{c}{$\chi_{cJ}\to\Xi^0\bar{\Xi}^0$} & \\
		\cline{2-4} \cline{6-8}
		Source &$\chi_{c0}$ &$\chi_{c1}$& $\chi_{c2}$ &&$\chi_{c0}$ &$\chi_{c1}$ &$\chi_{c2}$ & \\
		\hline
		Photon reconstruction        & 1.0 & 1.0 & 1.0 &  & 1.0 & 1.0 & 1.0 & \\
		$\Xi$ reconstruction         & 2.6 & 3.4 & 2.7 &  & 3.1 & 3.4 & 3.6 & \\
		$M_{\gamma\Xi}^{\rm recoil}$ mass window & 0.9 & 0.7 & 0.5 &  & 0.1 & 0.2 & 0.2 & \\
		Fitting range                & 0.6 & 0.5 & 0.7 &  & 0.5 & 1.2 & 0.8 & \\
		Signal shape                 & 1.6 & 3.2 & 4.2 &  & 3.5 & 5.9 & 6.2 & \\
		Background shape             & 0.2 & 0.3 & 0.2 &  & 1.6 & 1.2 & 1.1 & \\
		Angular distribution         & $-$ & 5.9 & 4.8 &  & $-$ & 2.1 & 3.9 & \\
		Branching fractions          & 2.2 & 2.6 & 2.2 &  & 2.2 & 2.6 & 2.2 & \\
		Total number of $\psi$(3686) & 0.6 & 0.6 & 0.6 &  & 0.6 & 0.6 & 0.6 & \\
		Total                        & 4.1 & 8.1 & 7.4 &  & 5.6 & 7.9 & 8.6 & \\
		\hline
	\end{tabular}
	\caption{\label{bfsys} Systematic uncertainties on the branching fraction measurements (in \%).}
\end{table}

\begin{table}[bpt]
	\centering
	\begin{tabular}{|ccccccccc|}
		\hline
		& \multicolumn{3}{c}{$\chi_{cJ}\to\Xi^-\bar{\Xi}^+$} && \multicolumn{3}{c}{$\chi_{cJ}\to\Xi^0\bar{\Xi}^0$} & \\
		\cline{2-4} \cline{6-8}
		Source &$\chi_{c0}$ &$\chi_{c1}$& $\chi_{c2}$ &&$\chi_{c0}$ &$\chi_{c1}$ &$\chi_{c2}$ & \\
		\hline
		Bin size               & 0.01 & 0.09 & 0.01 &  & 0.01 & 0.28 & 0.05 & \\
		$\cos\theta$ range     & 0.16 & 0.46 & 0.28 &  & 0.34 & 0.31 & 0.18 & \\
		Fitting range          & 0.01 & 0.01 & 0.02 &  & 0.01 & 0.03 & 0.02 & \\
		Signal shape           & 0.01 & 0.02 & 0.01 &  & 0.00 & 0.01 & 0.01 & \\
		Background shape       & 0.03 & 0.05 & 0.01 &  & 0.00 & 0.00 & 0.01 & \\
		Efficiency correction  & 0.05 & 0.06 & 0.10 &  & 0.11 & 0.08 & 0.11 & \\
		Total                  & 0.17 & 0.48 & 0.30 &  & 0.36 & 0.43 & 0.22 & \\
		\hline
	\end{tabular}
	\caption{\label{adsys} Systematic uncertainties on the angular-distribution measurements (absolute value).}
\end{table}

Several sources of systematic uncertainties in the branching fraction and angular distribution measurements are studied and summarized in table~\ref{bfsys} and \ref{adsys}, respectively. While some common systematic uncertainties of the BESIII experiment are briefly mentioned below and described elsewhere, the other specific systematic uncertainties are described in detail in the following paragraphs.

Systematic uncertainties associated with the branching fraction measurements include the photon reconstruction efficiency, the $\Xi$ reconstruction efficiency, the requirement on $M_{\gamma\Xi}^{\rm recoil}$, the fit procedure and the angular distribution. Other sources of systematic uncertainties arise from the total number of $\psi(3686)$ decays~\cite{DATASET} and the cited branching fractions~\cite{PDG}. The uncertainty of the photon reconstruction efficiency is estimated to be 1.0\% per photon~\cite{photon}. The systematic uncertainty for the $\Xi^-$ and $\Xi^0$ reconstruction efficiencies, including the tracking and PID efficiencies, $\pi^0$ reconstruction efficiency in the $\Xi^0$ decay, $\Lambda$ reconstruction efficiency as well as the requirements on mass window and decay length are studied using the methods  described in refs.~\cite{exp2, adpaper2, singletag0, BESIII:2021gca}. The uncertainty associated with the requirements on $M_{\gamma\Xi}^{\rm recoil}$ are estimated by smearing the $M_{\gamma\Xi}^{\rm recoil}$ distribution of MC sample according to the resolution difference between data and MC~\cite{masswin}. The systematic uncertainty due to the fit of the $M^{\rm recoil}_{\gamma}$ spectrum includes considerations of the fit range, signal shape and background shape. The systematic uncertainty associated with the fit range is estimated by varying the mass range in steps of 50 MeV/$c^{2}$. The systematic uncertainty due to the signal shape is estimated by changing the nominal signal shape to an individual simulated shape. To estimate the E1 transition effects~\cite{E1theory} on the signal shape, the correction method described in ref.~\cite{E1exp} is applied and the differences in the branching fractions are assigned as the systematic uncertainties. For the uncertainty due to the background shape, since the background is smoothly distributed in the region of interest, an alternative fit is performed using a third-order polynomial. The systematic uncertainty resulting from the angular distribution of the $\Xi$ baryon is estimated by the differences in branching fractions when alternative MC samples generated with $\alpha = 0$ are used. The total systematic uncertainty is obtained by summing the individual contributions in quadrature.

Systematic uncertainties associated with the angular-distribution measurements include the bin size and fit range of cos$\theta$, the fit procedure and efficiency correction in the determination of signal yields in each $\cos \theta$ interval. The systematic uncertainty associated with the bin size in $\cos\theta$ is estimated by changing the number of bins from 16 to 8 for the $\chi_{cJ}\to\Xi^-\bar{\Xi}^+$ process and from 10 to 5 for the $\chi_{cJ}\to\Xi^0\bar{\Xi}^0$ process. The systematic uncertainty due to the fit range of $\cos \theta$ is estimated by changing the $\cos \theta$ range from [$-0.8$, 0.8] to [$-0.7$, 0.7] for the $\chi_{cJ}\to\Xi^-\bar{\Xi}^+$ process and from [$-0.8$, 0.8] to [$-0.64$, 0.64] for the $\chi_{cJ}\to\Xi^0\bar{\Xi}^0$ process. The systematic uncertainty due to the fit of the $M^{\rm recoil}_{\gamma}$ spectrum includes considerations of the fit range, signal shape and background shape. These are estimated using the same method as described above. In this analysis, the $\alpha$ parameters are determined by fitting the efficiency-corrected $\cos \theta$ distribution. The systematic uncertainty associated with the efficiency used to perform the correction is estimated by changing the MC samples generated with $\alpha ~ (\alpha') = 0$ to the measured $\alpha ~ (\alpha')$ value. The total systematic uncertainty is obtained by summing the individual contributions in quadrature.

\section{Summary}
The branching fractions and angular distributions for $\chi_{cJ}\to\Xi^-\bar{\Xi}^+$ and $\Xi^0\bar{\Xi}^0$ processes are measured using $448.1 \times 10^6$ $\psi(3686)$ decays collected with the BESIII detector, 
The results show that the branching fraction of the $\chi_{cJ}\to\Xi^-\bar{\Xi}^+$ process is consistent with that of the  $\chi_{cJ}\to\Xi^0\bar{\Xi}^0$ process within one or two standard deviation, as expected from isospin symmetry. The decays $\chi_{c1}\to\Xi^0\bar{\Xi}^0$ and $\chi_{c2}\to\Xi^0\bar{\Xi}^0$ are observed for the first time with statistical significances of 7$\sigma$ and 15$\sigma$, respectively. The measured branching fractions in this analysis are still inconsistent with theoretical predictions~\cite{theory1, com1, com3}. On the contrary, the measured angular distributions of $\chi_{c0}\to\Xi^-\bar{\Xi}^+$ and $\Xi^0\bar{\Xi}^0$ processes are consistent with theoretical expectations~\cite{helicity_amp}.  The precision for these measurements are significantly improved compared to the previous measurements~\cite{exp3, bes2_xixibar}.

\acknowledgments
The BESIII collaboration thanks the staff of BEPCII, the IHEP computing center and the supercomputing center of USTC for their strong support. This work is supported in part by National Key R\&D Program of China under Contracts Nos. 2020YFA0406400, 2020YFA0406300; National Natural Science Foundation of China (NSFC) under Contracts Nos. 11335008, 11625523, 11635010, 11735014, 11705192, 11822506, 11835012, 11905236, 11935015, 11935016, 11935018, 11961141012, 11950410506, 12022510, 12025502, 12035009, 12035013, 12047501, 12075107, 12061131003, 12061131003, 12122509, 12105276; the Chinese Academy of Sciences (CAS) Large-Scale Scientific Facility Program; Joint Large-Scale Scientific Facility Funds of the NSFC and CAS under Contracts Nos. U1732263, U1832207, U1832103, U2032111; CAS Key Research Program of Frontier Sciences under Contract No. QYZDJ-SSW-SLH040; 100 Talents Program of CAS; INPAC and Shanghai Key Laboratory for Particle Physics and Cosmology; Fundamental Research Funds for the Central Universities grant No. WK2030000053; ERC under Contract No. 758462; European Union Horizon 2020 research and innovation programme under Marie Sklodowska-Curie grant agreement No. 894790; German Research Foundation DFG under Contracts Nos. 443159800, Collaborative Research Center CRC 1044, FOR 2359, GRK 214; Istituto Nazionale di Fisica Nucleare, Italy; Ministry of Development of Turkey under Contract No. DPT2006K-120470; National Science and Technology fund; Olle Engkvist Foundation under Contract No. 200-0605; STFC (United Kingdom); The Knut and Alice Wallenberg Foundation (Sweden) under Contract No. 2016.0157; The Royal Society, UK under Contracts Nos. DH140054, DH160214; The Swedish Research Council; U. S. Department of Energy under Contracts Nos. DE-FG02-05ER41374, DE-SC-0012069.

\end{document}